\documentclass[aps,pre,two column]{revtex4-2}
\usepackage[utf8]{inputenc}
\usepackage{amsmath}
\usepackage{amsfonts}
\usepackage{graphicx}
\usepackage{subfig}
\usepackage{xcolor}
\usepackage{csquotes}
\usepackage{gensymb}
\usepackage{mathrsfs}
\usepackage{csquotes}
\usepackage{natbib}

\begin{document}
	\title{Condensate-induced organization of the mass profile and emergent power laws in the Takayasu aggregation model  }
	\author{Reya Negi, Rajiv G Pereira, and Mustansir Barma}
	\affiliation{TIFR Centre for Interdisciplinary Sciences, Tata Institute of Fundamental Research, Gopanpally, Hyderabad, 500046, India}
	\begin{abstract}
		Large-mass condensates, which coexist with a power-law-decaying distribution in the one-dimensional Takayasu model of mass aggregation with input, were recently found in numerical simulations. Here, we establish the occurrence of condensates by analyzing exact recursions for finite systems and further show that they have a strong effect on the properties of the system. In the steady state of a large but finite system, there is a single condensate, whose \textcolor{black}{random} movement through the system leads to a reorganization of the mass profile on a macroscopic scale. A scaling analysis of the mean mass and standard deviation at different distances from the condensate leads to the surprising conclusion that the mass distribution on a macroscopic number of sites \textcolor{black}{around the} condensate follows a power-law decay with an exponent 5/3, while farther-away sites show the customary Takayasu exponent 4/3, with a crossover in between. Finally, the exit of condensates from a system with open boundaries has a strong effect on the temporal fluctuations of the total mass in the steady state. Their departure is followed by a buildup of mass and subsequent departures, leading to strong intermittency, established through a divergence of the flatness as the scaled time approaches zero. 
	\end{abstract}

	\maketitle
	\section{Introduction}
	Aggregation phenomena are important in many areas of the natural sciences, cutting across different fields and length scales, from clouds and river networks on the one hand, to colloid
	formation and biomolecular organization on the other \cite{friedlander2000,scheidegger1967,smoluchowski1917,brangwynne_2011,brangwynne_2018,brangwynne_2023,arsenadze2024anomalous}. The subject, which began with the pioneering studies of colloidal clusters by Smoluchowski \cite{smoluchowski1917}, continues to find new domains of applicability, most recently in aspects of biomolecular organization \cite{brangwynne_2011,arsenadze2024anomalous,brangwynne_2018,brangwynne_2023}.
	
	The kinetics of aggregation involves the coalescence and breakup of clusters with different numbers of particles and, thus different masses. The resulting time-dependent mass distribution often shows interesting features. First, the distribution may be broad, characterized by a power law, signaling scale invariance arising from critical correlations in the
	system. Secondly, spatially well-separated condensates may form, each one being a single cluster that holds a very large number of constituent particles. It is natural to ask the following questions: Under
	what conditions do these features manifest$?$ Can they arise together, and if so, under what
	circumstances$?$
	
	Much can be learned from the analysis of simple models whose definitions include only the essential kinetic processes, but whose emergent behavior is nontrivial. The Takayasu
	model of aggregation is a prime example \cite{takayasu_1988,privman,Takayasu_1991}. Its moves include diffusion of particle clusters, followed by coalescence when two clusters meet. Further, there is an input of
	particles at a steady rate. The appeal of the model stems from its simplicity and the fact that as time passes, the distribution of particle numbers approaches a power-law scaling function
	of mass and time, indicating a self-organized critical (SOC) state. This can be established analytically within mean field theory and through an exact recursive calculation in 1D \cite{takayasu_1988,Takayasu_1989,privman,Huber_1991}. In addition, spatiotemporal correlation functions can be calculated exactly \cite{SN_2000}. Interestingly, the Takayasu model has exact correspondences to several other models, for instance the Scheidegger model of river basins \cite{scheidegger1967,SN_2000}, the voter model \cite{liggett2013} and models of stress propagation in bead packs \cite{coppersmith1996}.
	Further, experiments indicate that nucleoli in certain living cells share the Takayasu moves of diffusion, coalescence and input, and show a broad power-law distribution of sizes \cite{brangwynne_2011,brangwynne_2018,brangwynne_2023}.
	
	Whereas the power law scaling is well established and known since the inception of the
	Takayasu model, it is only recently that it was realized that the model also exhibits condensates or extremely large-mass clusters \cite{Arghya_2023} \textcolor{black}{at both intermediate and long times.} Condensates are known to arise in
	aggregation models with mass conservation, where their formation is akin to Bose-Einstein condensation \cite{evans2005,majumdar1998nonequilibrium,majumdar2000nonequilibrium,das2015additivity,iyer2023}. But in the Takayasu model, where the mass is strongly non-conserved owing to the steady input, the mechanism is quite different. \textcolor{black}{It is an effect arising from the influence of the maximal term in slowly decaying power-law distributions~\cite{darling,feller1991introduction}. Numerical simulations show that due to this maximal term effect,} beyond the power law, the probability distribution of masses in the Takayasu model exhibits a strong and distinctive
	condensate hump, which is part of the scaling function \cite{Arghya_2023}, confirmed by the numerically exact recursive results reported below. A time-dependent length scale $\mathcal{L}(t)$ determines condensate masses and the spacing between them. Since the underlying process is diffusive, $\mathcal{L}(t)$ is proportional to $t^{1/2}$. This implies that as $t$ increases, the number of condensates falls while the size of each increases, as in phase ordering dynamics \cite{iyer2023,bray1994theory}. This continues until $\mathcal{L}(t)$ reaches the system size $L$, beyond which a long time state with statistically invariant properties sets in \cite{Arghya_2023}.
	 \begin{figure*}
		\subfloat[\label{fig:recurssion1}]{\raisebox{0mm}{
				\includegraphics[width=0.71\columnwidth]{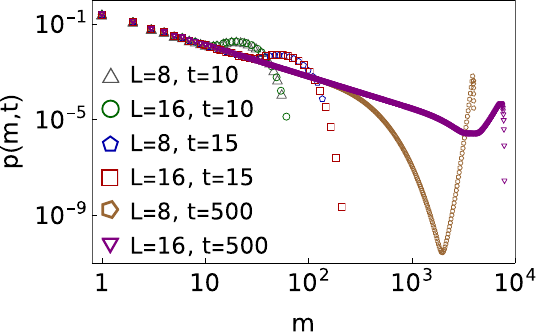}}
		}
		\subfloat[\label{fig:scalingcor}]{\hspace{-3mm}{
				\includegraphics[width=0.67\columnwidth]{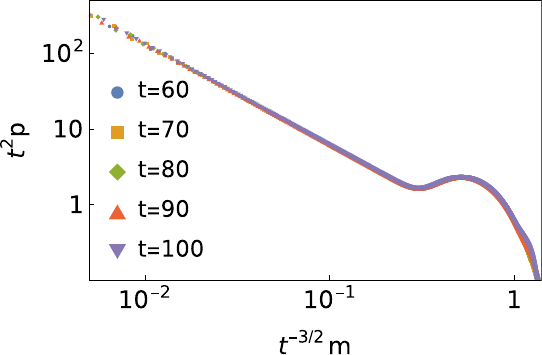}}
		}
		\subfloat[\label{fig:scalingss}]{\raisebox{0mm}{
				\includegraphics[width=0.64\columnwidth]{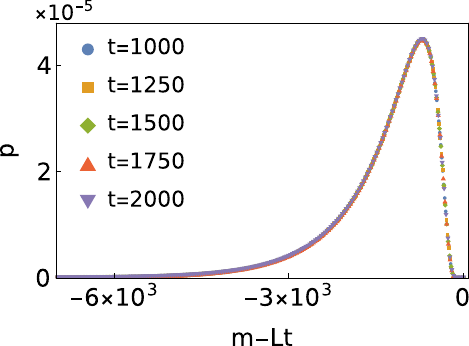}
		}}
		\caption{(a): The probability distribution $p(m, t)$  for $L=8$ and $16$ in the regimes $t \ll t^*=L^2$ and $t \gg t^*$. In the former regime ($t=10$ and $15$)  the condensates correspond to the hump, while in steady state ($t=500$), there is a single condensate corresponding to the peak \textcolor{black}{near} the right edge. (b) and (c): Scaling of $p(m,t)$ in the regimes $t \ll t^*$ and $t \gg t^*$, respectively, for $L=16$. }
		\label{fig:recurssion2}
	\end{figure*}
	
	Condensates have a strong effect on the properties of the Takayasu model. In this paper, we first establish their occurrence \textcolor{black}{in finite systems} by an exact calculation of the probability distribution of mass as a function of time (see Fig. \ref{fig:recurssion1}) through recurrence relations of the mass held in a stretch of lattice sites
	\cite{takayasu_1988,privman,Huber_1991}. Further,  we investigate the effect of condensates on the mass profile in the system, recognizing that a condensate functions like an efficient forager which ingests mass from neighboring regions. The result is surprising: with periodic boundary conditions, the \textcolor{black}{mean} mass profile seen
	by the condensate is \textcolor{black}{strongly influenced} on a \textit{macroscopic} scale. The mean and
	standard deviation are both scaling functions of $r/L$ where $r$ is the distance from the
	condensate. As shown below, the results imply that the distribution of masses on sites close to the condensate follows a new power law $\sim m^{-5/3}$ as opposed to the well-known Takayasu
	power $\sim m^{ -4/3}$, which holds on distant sites. Finally, we investigate the time variation of the mass of a system with free boundary conditions and find that the build-up and departure of condensates leads to enormous fluctuations of the total mass, characterized by strong intermittency, reminiscent of dragon king events \cite{sornette2012dragon,rdsouza}.

	\section{The Model and earlier work}

	In $1$D, the Takayasu model is defined on a lattice with $L$ sites with site $i \in \{1,\;2,..L\}$ holding a mass $m(i,t)$, which is a whole number. We assume periodic boundary conditions so that $m(i,t) = m(i+L,t)$.  
	At each time step the following two steps occur in the system. (1)~Diffusion: the entire mass at each site either hops to the neighboring site to the right, or stays put, with probability $1/2$ each. Note that this can be mapped to the familiar symmetric diffusion of a random walker on moving  to a frame that has a velocity $1/2$ with respect to the lattice~\cite{Takayasu_1989}. Evidently, the mass distribution is the same in either description. (2)~Input: a unit mass is injected at each site.
	
	The dynamics of the system are thus described by the stochastic equation
	\begin{equation}\label{stochasticEq}
		m(i,t+1) = (1-W_i(t)) m(i,t) + W_{i-1}(t) m(i-1,t) + I_i(t),
	\end{equation}
	where $W_i(t)=0$ or $1$ with probability $1/2$ each describes mass movement. \textcolor{black}{The random variable $I_i(t)$ describes the process of injection. With stochastic insertion, $I_i(t)$ is $1$ or $0$ depending on whether or not unit mass is injected on site $i$ at time $t$. We consider the case of definite injection at every site at every time step, implying $I_i(t)=1$. }  
	
	\textcolor{black}{Earlier, it was found that in the limit $L\rightarrow \infty$, followed by  $t\rightarrow \infty$, the mass distribution follows a power-law decay $p(m) \sim m^{-4/3}$ \cite{Takayasu_1989}. If $t$ is finite, there is a time-dependent length proportional to $ t^{1/2}$ within which masses are correlated and beyond which they are not. An exact calculation shows that the equal-time two-point correlation function follows a scaling form  $\langle m(i,t) m(i+r,t)\rangle_c=t^2f(r/\mathcal{L}(t))$,  with $\mathcal{L}(t)= t^{1/2}$, in the limit $r\rightarrow \infty,t\rightarrow \infty$ with $r/\mathcal{L}(t)$ finite \cite{SN_2000}.}
		
		\textcolor{black}{Recent numerical simulations have revealed the presence of condensates in the system \cite{Arghya_2023}, which lead to a characteristic hump in the mass distribution at finite times. The condensate accumulates a significant portion ($\sim 80\%$) of the total mass in the system, and in the long-time limit, the system contains a single condensate with mass of order $O(L^{3})$. }
	
	\section{The maximal term and condensates}
	
	The largest of a number of terms drawn independently from \textcolor{black}{a class} of L\'{e}vy \textcolor{black}{distributions} has the attributes of a condensate, in that on average, the ratio of the sum of all the terms to the largest term is finite.
	
	Consider $n$ random variables $m_1,m_2,....m_n$. Suppose they are independent and identically distributed with the probability distribution function $p(m) \sim m^{-(1+\alpha)}$ with $\alpha<1$. Let $S_n$ and $M_n$ denote their sum and maximum respectively:
	$S_n=\sum_{i=1}^{n} m_i$ and $M_n = \max(m_1, m_2,..., m_n)$. The variables $S_n$ and $M_n$ are strongly correlated and satisfy \cite{darling,feller1991introduction}
	\begin{equation}\label{fellar}
		\langle S_n / M\\
		_n \rangle \rightarrow \frac{1}{1-\alpha},
	\end{equation}
	This result indicates that the maximal term carries a finite fraction of the total, which is the hallmark of a condensate.
	
	 Equation (\ref{fellar}) holds for independent, identically distributed variables, and the question arises whether a similar mechanism for condensate formation operates also when variables are correlated. For the Takayasu model, the answer is in the affirmative despite correlations between masses on different sites. It seems that the existence of the condensate rests on the occurrence of the \textcolor{black}{slow} power law $p(m) \sim m^{-4/3}$, while the correlations between masses have the relatively minor effect of changing the asymptotic value of $\langle S_n / M_n \rangle$ from $3/2$ (in the absence of correlations) to $\simeq 1.41$ (\textcolor{black}{obtained by numerically averaging $S_n/M_n$ in subsystems of size $\mathcal{L}(t)=t^{1/2}$  over many histories.)}
	
	\textcolor{black}{An interpretation of the condensate mass in terms of L\'{e}vy distributions is possible even if $\alpha>1$ \cite{majumdar2005}, in which case $S_n$ would be proportional to $n$. The interesting point here is that Eq. (\ref{fellar}) shows that the interpretation holds even if $\alpha<1$ and $S_n$ grows as $n^{1/\alpha}$.
	}
	
	\section{Exact solution}
	 Our objective is to calculate the probability $p(m,t)$ to find a mass $m$ at any site at time $t$, given an initial configuration with every site occupied by a unit mass.  

	Following~\cite{Takayasu_1989}, we consider the mass contained in a stretch of $r$ consecutive sites at any instant:
	\begin{align}\label{cluster}
		&M_{n,r}(t) = \sum_{i=1}^{r} m(n+i-1,t), &r=1,\;2,...L.
	\end{align}
	The corresponding probability distribution, which is translationally  invariant,  has a characteristic function    
	\begin{equation}\label{charFn}
		Z_{r}(q,t) \equiv \left\langle  \exp{\left\{iq M_{nr} (t)\right\}} \right\rangle,
	\end{equation}
	where the average is over all histories.
	Equations~(\ref{stochasticEq})~and~(\ref{cluster}), lead to the recursion relation~\cite{Takayasu_1989}
	\begin{align}\label{recurrence}
		Z_r(q, t + 1) = \frac{ \phi(q)^r}{4} \left[ Z_{r+1}(q, t) + Z_{r-1}(q, t) + 2 Z_r(q, t) \right],
	\end{align}
	with the initial condition $Z_r(q,0) = \exp{(iqr)}$, and the boundary conditions $Z_0(q,t) = 1$ and $Z_L(q, t) = \exp(iq L(t+1))$. \textcolor{black}{Here, the prefactor $\phi(q) \equiv \langle \exp \left\{i q I_i(t)\right\} \rangle$. Earlier, Takayasu et al.~\cite{Takayasu_1989, Takayasu_1991} and Huber~\cite{Huber_1991} obtained the $L=\infty$, $t = \infty$ mass distribution to leading order in $m$. They proceeded by setting $Z_r(q,t+1) = Z_r(q,t)$ in Eq.~(\ref{recurrence}) and expanding $\phi(q)$ as }
	\begin{equation}\label{expansionphi}
		\phi(q) = 1 + i \langle I \rangle q - \frac{1}{2} \langle I^2 \rangle q^2 + ...
	\end{equation}
	\textcolor{black}{The leading behavior of $p(m)$ with $m$ is obtained by retaining the first two non-vanishing terms in expansion~(\ref{expansionphi}): for random mass-injection, the leading behavior  is captured by the $\langle I \rangle$ term, yielding $p(m) \sim m^{-4/3}$.}
	
	We remark that the condensates did not appear in the treatments of~\cite{Takayasu_1989,Takayasu_1991, Huber_1991}, as they are missed in the limit $t \rightarrow \infty$ in an infinite system.
	We obtain  exact results for finite $L$ and $t$ below and show the existence of condensates, corroborating the conclusions of~\cite{Arghya_2023}, which were based on numerical simulations.
	
	To this end, we first iterate Eq.~(\ref{recurrence}) $t$ times for a given $L$, with the help of a computer, to obtain $Z_1 (q, t)$, and  then perform an inverse Fourier transform on $Z_1(q,t)$ yielding
	\begin{equation}
		p(m,t) = \frac{1}{2 \pi} \int_{-\pi}^{\pi} Z_{1}(q,t) e^{iqm}dq.
	\end{equation}
	This procedure leads to the exact distribution $p(m,t)$ for any $L$ and $t$.

	\textcolor{black}{In a system of size $L$, there is a characteristic time $t^* = L^2$ over which mass fluctuations spread diffusively across the system.} For $t \ll t^*$, fluctuations spread across a length $\mathcal{L}(t) = \sqrt{t} \ll L$, whereas for $t$ exceeding $t^*$, the system reaches a steady state in a sense to be made precise below.   Below, we discuss our results for $p(m,t)$ separately for the two regimes: (a)  $t \ll t^*$ and (b) $t \gg t^*$.
	\textcolor{black}{We will find that condensates show up in both the regimes --- as a hump in the former regime and as a sharper peak in the latter.}
	We restrict ourselves to relatively small values of $L$, namely, $L=8$ and $16$, as they suffice to show all the relevant features of $p(m,t)$ and as going to higher values of $L$ is computationally costly.

	(a) To study $p(m,t)$ in the regime $t \ll t^*$, we examine the cases $t=10$ and $15$ for $L=8$ and $16$ shown in Fig.~\ref{fig:recurssion1}: the distribution is characterized by the power-law $p(m) \sim m^{-4/3}$ followed by a hump. The hump corresponds to the presence of large condensates in the system, with typically one condensate in each region of length $\mathcal{L}(t)$~\cite{Arghya_2023}. The distribution is cutoff beyond the hump at $m=(t+1)L$, which is the total mass in the system at time $t$. For both values of $L$, the distributions $p(m,t)$  are identical except for the cutoff, \textcolor{black}{a property that we expect will continue to hold for any $L$.}
	For different values of $t$, on scaling $m$ by $t^{-3/2}$ and $p(m,t)$ by $t^2$, for $t \ll t^*$, the curves collapse (see~Fig. \ref{fig:scalingcor}), implying the scaling form $p(m,t) \approx t^{-2} f(u)$, where $u=mt^{-3/2}$. Note that the condensate hump is part of the scaling function. The region $u<u_0\simeq 0.6$ shows the aforementioned power-law decay, while the condensate hump appears for $u>u_0$, with $f(u)$ decaying exponentially as $u \rightarrow \infty$~\cite{Arghya_2023}.      
	
\begin{figure*}
		\subfloat[\label{fig:pOfmsup}]
		{
			\raisebox{0cm}
			{\includegraphics[width=0.75\columnwidth]{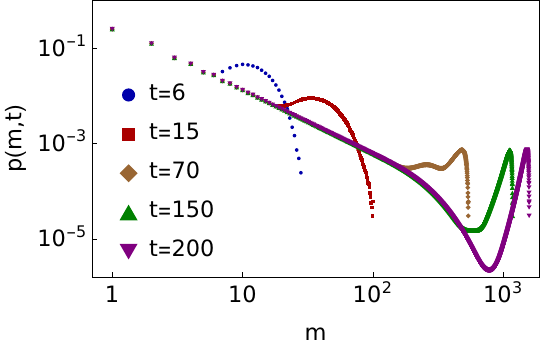}}
		}
		\subfloat[\label{fig:z1t6sup}]{
			\includegraphics[width=0.7\columnwidth]{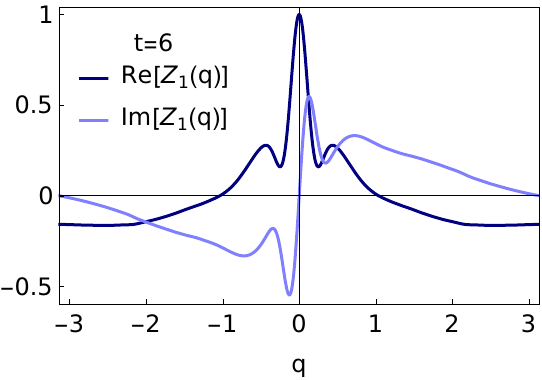}
		}
		
		\subfloat[\label{fig:z1t15sup}]{
			\includegraphics[width=0.67\columnwidth]{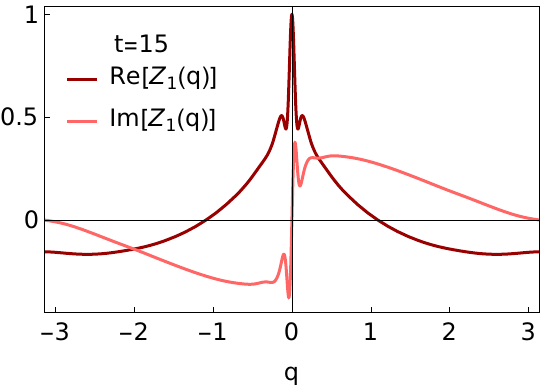}
		}
		\subfloat[\label{fig:z1t100sup}]{
			\includegraphics[width=0.67\columnwidth]{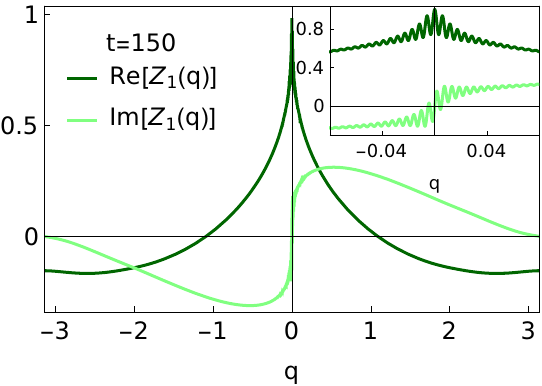}
		}
		\subfloat[\label{fig:z1t200sup}]{
			\includegraphics[width=0.67\columnwidth]{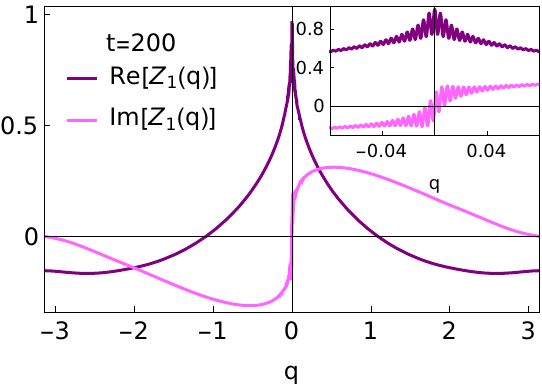}
		}
		\caption{
			(a): Evolution of the probability density profile for $L=8$. (b)-(e) The Fourier transform $Z_1(q)$ at different instants in the regimes $t \ll t^*$ and $t \gg t^*$ for $L=8$. For  $t=6$ and $15 \ll t^*$, the condensates, which lead to the hump in $p(m,t)$, correspond to  secondary maxima near the origin in $Z_1(q)$. In steady state ($t=150$ and $200$), the condensates, which appear as a peak at the right edge in $p(m,t)$, correspond to rapid oscillations near the origin in $Z_1(q)$ .
		}
		\label{fig:recurssion}
	\end{figure*}

	The power-law $m^{-4/3}$, is reflected by the cusp (see Figs.~\ref{fig:z1t6sup} and \ref{fig:z1t15sup}) in $Z_1(q,t)$, which varies as $q^{1/3}$. The secondary maxima near the origin in the real and the imaginary parts of $Z_1$, shown in Figs.~\ref{fig:z1t6sup} and \ref{fig:z1t15sup} for $t=6$ and $15$, respectively, correspond to the condensate hump. As time increases from $t=6$ to $15$, the condensate hump shifts to larger values of $m$, and the secondary maxima in $Z_1(q,t)$ move closer to the origin.
	
	(b) For a fixed $t$ in the regime $t\gg t^*$,  the distribution follows $p(m,t) \sim m^{-4/3}$ for $m\ll m^*$ and $p(m) \sim \exp(-m/m^*)$ for $m \gg m^*$, where $m^*= L^3 +L$ is a characteristic mass equal to the total mass in the system up to time $t^*$.  Toward the end of this exponential tail the distribution exhibits a prominent peak, representing a single large condensate [see Fig. \ref{fig:recurssion1}, L = 8, 16; t = 500]. The area under the peak is $1/L$, the probability to find the condensate at a randomly chosen site.
	In the steady state, the condensate peak in the distribution  is seen to follow $p(m,t) \approx H(m-Lt)$ (see Fig. \ref{fig:scalingss}), showing that once $t$ exceeds $t^*$, practically all the mass injected into the system is eventually absorbed by the condensate. Concomitantly, the mass distribution of the rest of the system, which corresponds to the portion before the condensate peak, assumes a time-independent form. Thus, we shall refer to the $t \gg t^*$ regime as \enquote*{steady state}, despite the fact that the condensate mass keeps growing.  
	
	Figures~\ref{fig:z1t100sup} and \ref{fig:z1t200sup} show $Z_1(q,t)$ after the steady state has been reached, at $t=150$ and $200$, respectively, for $L=8$.  As before, the cusps here correspond to the power-law decay $p(m) \sim m^{-4/3}$~\textcolor{black}{(see Fig.~\ref{fig:pOfmsup})}. The condensate part, which is a sharp peak in the large mass region, is reflected as rapid oscillations near the origin in $Z_1(q,t)$, reminiscent of the Fourier transform of a delta function $\delta(m-m_c(t))$, where $m_c$ is the mass of the condensate. As time increases from  $t=150$ to $200$, the condensate mass $m_c(t)$ increases, and the frequency of oscillations in $Z_1(q,t)$ increases.


	\section{Mass profile with respect to the condensate}
	
	The condensate moves around the system accumulating mass, which raises intriguing questions about its impact on the surroundings. To address this, we perform numerical simulations and analyze the time-averaged mass profile \textit{relative to the condensate}. We monitor the mean and standard deviation of the masses at position $r$ with respect to the condensate in a system of size $L$. In other words, we shift the origin to the position of the condensate at each time step.
	\textcolor{black}{Remarkably, we find that the properties of the system in the frame co-moving with the condensate are statistically stationary. }
	\begin{figure}
		\subfloat[\label{subfig:mean mass}]{
			\includegraphics[width=0.89\columnwidth]{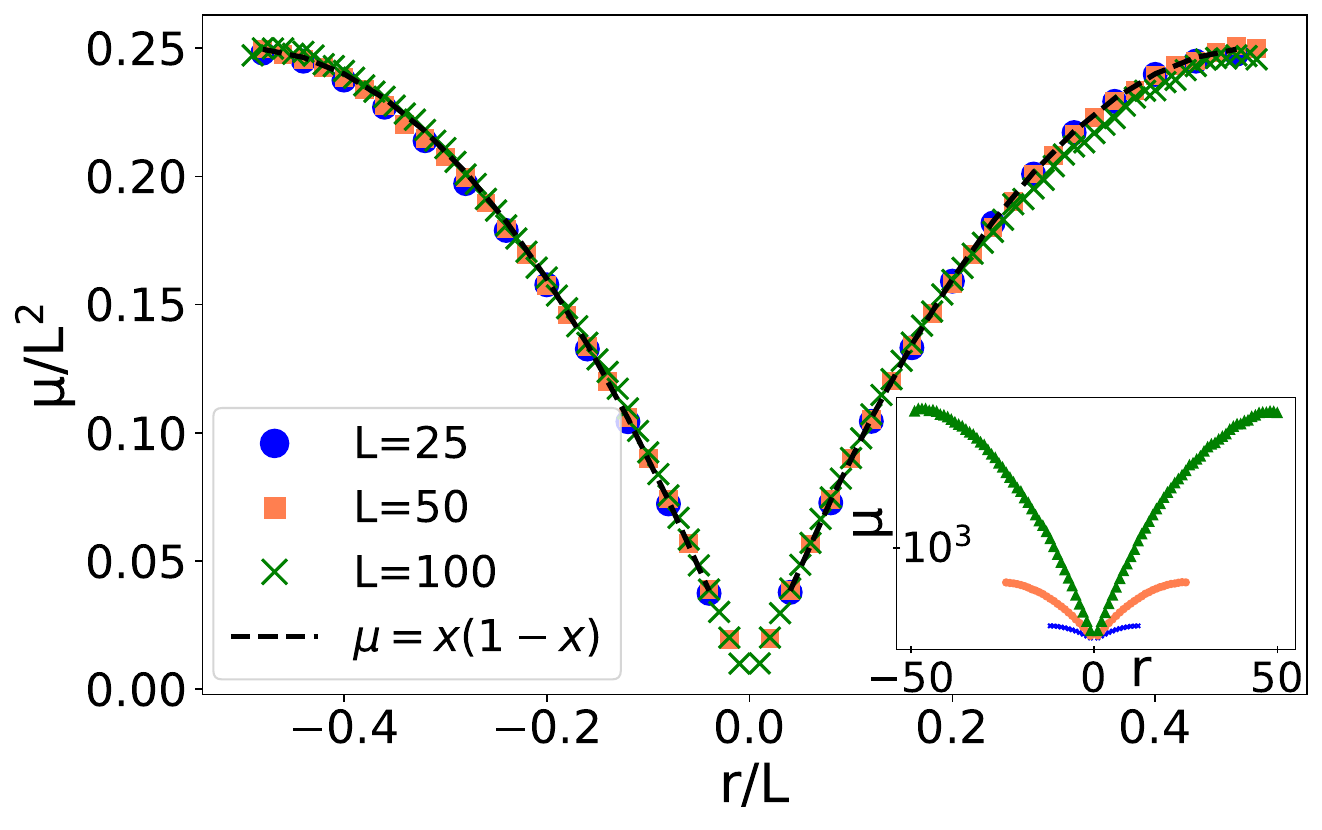}
		}
		
		\subfloat[\label{subfig:std}]{
			\includegraphics[width=0.89\columnwidth]{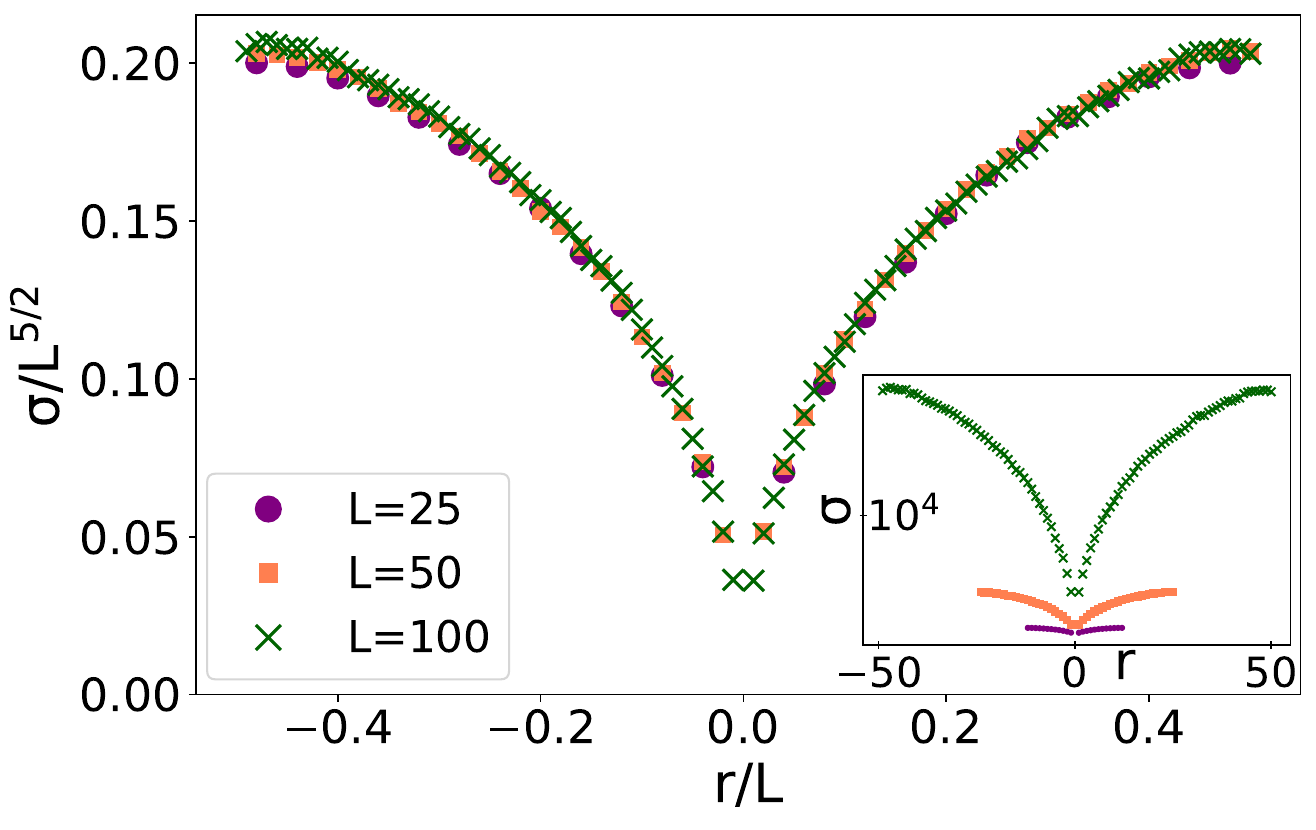}	
		}
		\caption{(a) Scaling of $\mu$ with $L$.   (b) Scaling of $\sigma$ with $L$.    The plots are for $L=25, \; 50$ and $100$. The insets show the unscaled plots.}
		\label{}
	\end{figure}
	
	Figure \ref{subfig:mean mass} shows that in steady state, the mass profile across the full system is strongly affected by the moving condensate. The main plot depicts scaling of the mean mass $\mu(r,L) = L^2 g_1\left(\frac{r}{L}\right)$. Our data is consistent with $g_1(x) = x(1-x)$, a form that we establish analytically. Similarly, in Fig. \ref{subfig:std}, the main plot shows the scaling of the standard deviation $\sigma (r,L) = L^{5/2}  g_2\left(\frac{r}{L}\right)$. We observe that $g_2(x)\sim x^{1/2}$ when $x\ll 1.$
	
	\subsection{Analytic form for $\mu(r,L)$}
	Consider a system of size $L$ with periodic boundary conditions following the Takayasu dynamics (see Eq.(1)).  Let $\mu(r,L)$ and $J(r,L)$ denote the mean mass and the mean mass current, respectively, at position $r$ with respect to the location of the condensate at steady state. As the masses  follow symmetric diffusive dynamics, we  write
	\begin{equation}
		J(r,L) = -D[\mu(r,L)-\mu(r-1,L)],
	\end{equation}
	where $D$ is the diffusion constant and the lattice spacing is $1$. In the continuum limit this reduces to
	\begin{equation}
		J=-D \partial \mu/\partial r.
		\label{eq2}
	\end{equation} 
	Evidently, in steady state, $J(r,L)$ must satisfy the continuity equation with injection
	
	\begin{equation}
		\partial J/\partial r = w,
		\label{eq1}
	\end{equation}
	where $w$ is the mass injected per unit length per unit time.
	Using Eqs.~(\ref{eq2}), ~(\ref{eq1}), and the scaling form  $\mu(r,L) = L^2 g_1\left(x\right)$, where $x=r/L$, we obtain the diffusion equation
	\begin{equation}
		D \frac{\partial^2 g_1}{\partial x^2}  = -w,
		\label{eq3}
	\end{equation}
	with the boundary conditions $g_1(0)=0$ and $g_1'(1/2)=0$. The former is obtained by considering the condensate as an absorbing boundary and the latter follows as $\mu$ flattens as $x \rightarrow 1/2$ corresponding to a local maximum. Equation~(\ref{eq3}) is readily solved to obtain 
	\begin{equation}
		g_1(x)=x(1-x),
	\end{equation}
	with $w/2D=1$ which matches the data.
	
		\begin{figure}
		\subfloat[\label{subfig:steady state}]{
			\includegraphics[width=0.88\columnwidth]{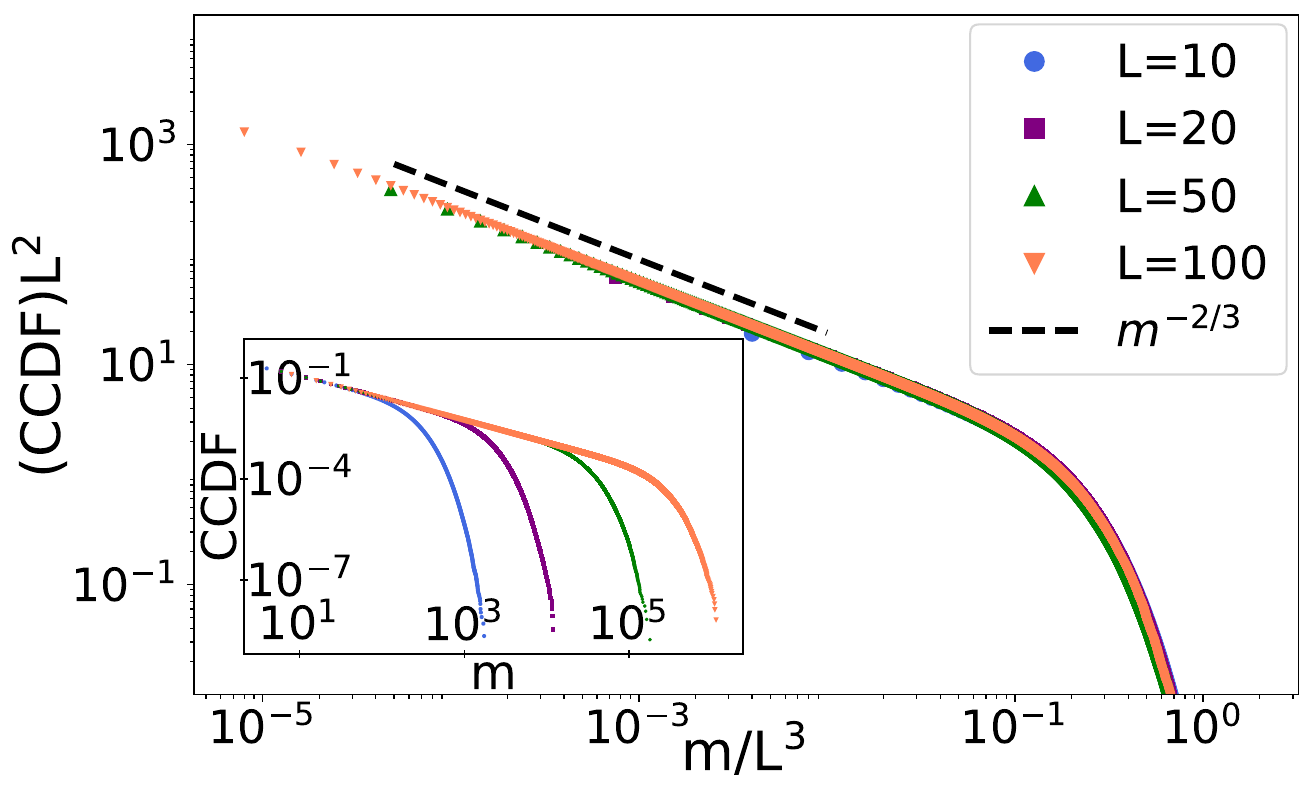}
		}
		
		\subfloat[\label{subfig:cumulative}]{
			\includegraphics[width=0.9\columnwidth]{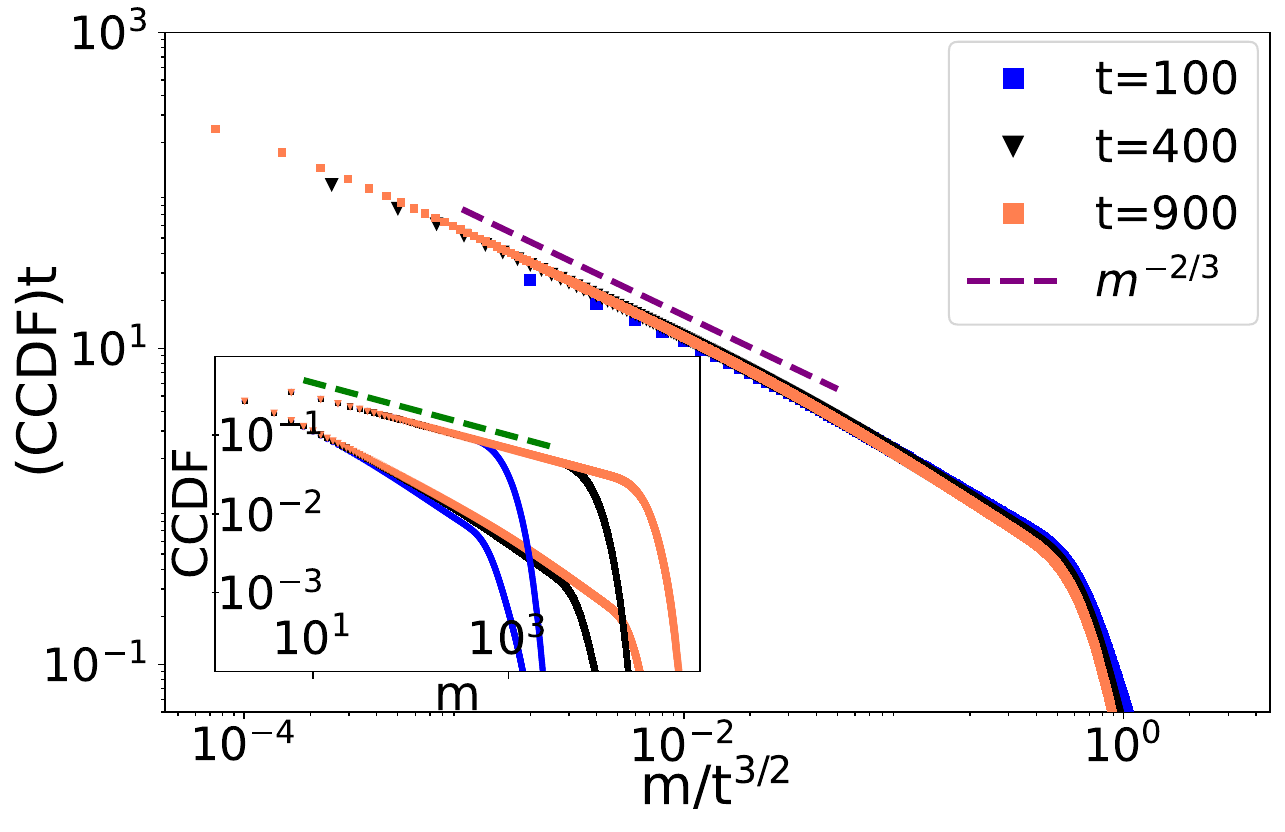}	
		}
		\caption{(a) Scaling of the complementary cumulative distribution function (CCDF) for the masses on the sites next to the condensate with $L$  for $L=10$, $20$, $50$, and, $100$ at steady state. Inset: Unscaled plots. (b) Scaling of the CCDF for masses on the sites next to the condensate with $t$ in the regime $t \ll t^*$ for $L=1200$. The inset shows CCDF vs $m$ plotted at different times for masses on the sites next to the condensate and masses in the bulk along with the lines  $m^{-2/3}$ and $m^{-1/3}$, respectively.}
		\label{}
	\end{figure}
	The scaling forms for the mean and the standard deviation imply that for a given $L$ the mass distribution $P(m)$ varies with $r$ and that it is a function of the scaled separation \( r/L \). We observe that in the regime $r/L \lesssim 1/2$, the behavior of $\mu$ and $\sigma$ are consistent with the power-law  
	\begin{equation}\label{theta}
		P(m) ~ \sim m^{-\theta}
	\end{equation}
	with $\theta = 4/3$. For a distribution of the form~(\ref{theta}) with $\theta<2$, we obtain the dependences \( \mu \sim L^{6-3\theta} \) and \( \sigma \sim L^{(9-3\theta)/2} \). For \( \theta = 4/3 \) this yields \( \mu \sim L^2 \) and \( \sigma \sim L^{5/2} \), matching the dependences obtained above from the scaling functions $g_1$ and $g_2$ for  $r/L \lesssim 1/2$.
	
		\begin{figure}
		\centering
		\includegraphics[width=\columnwidth]{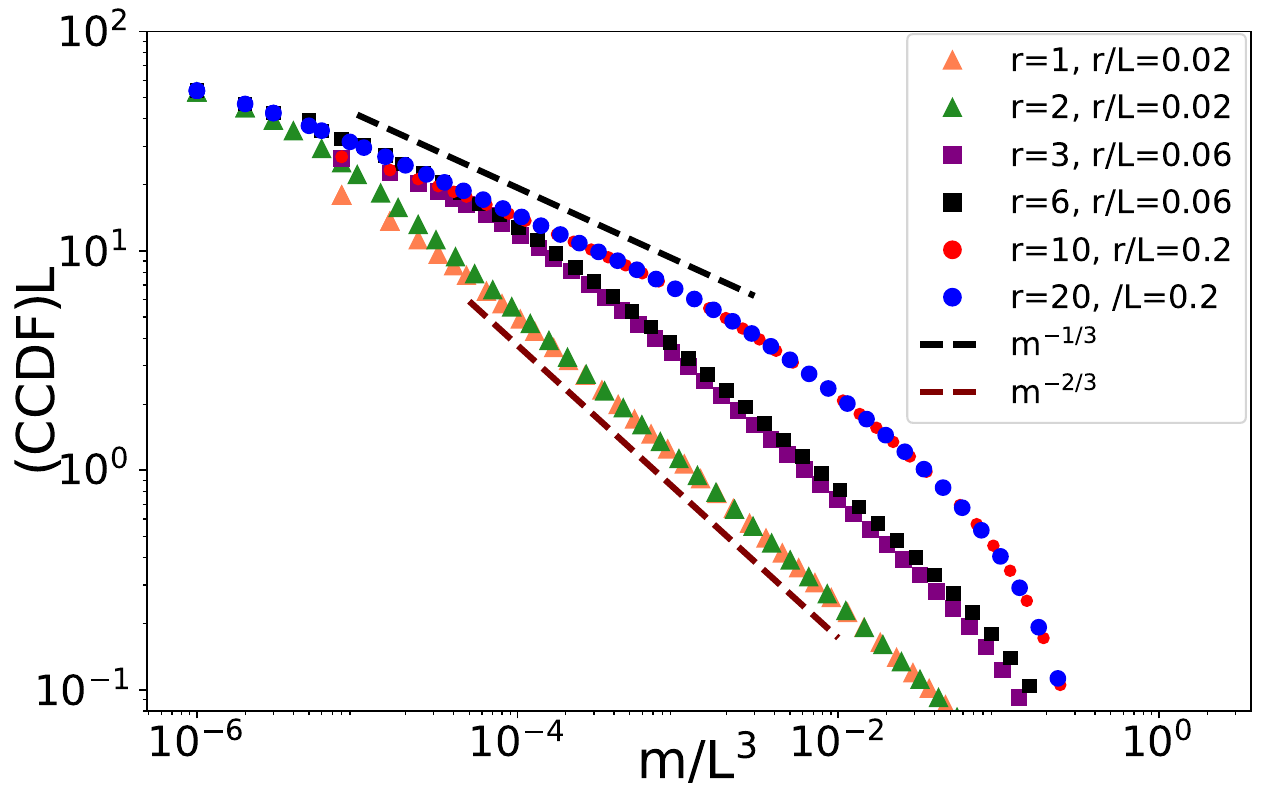}
		\caption{The scaled plot of CCDF vs $m$ for system sizes $L=50$ and $100$. We see that CCDF$\sim m^{1-\theta}$ with  $\theta=5/3$ for $r/L\ll 1$ and $\theta=4/3$ for $r/L \lesssim 1/2$, depicting a clear crossover.}
		\label{fig:crossover}
	\end{figure}
	
	However, close to the condensate, \textcolor{black}{where \( r/L \) is small $(\lesssim 0.1)$}, the moments $\mu$ and $\sigma$ scale differently, with \( \mu \sim L \) and \( \sigma \sim L^2 \), implying a new value of power-law decay exponent $\theta =5/3$. Thus, the value of the exponent changes from \( \theta = 4/3\) to \( \theta = 5/3 \) as one transitions from the region far from the condensate  $r/L \sim 1/2$ to the region close to it, highlighting a distinct change of the mass distribution.

	To check this directly, we monitor the mass distribution at sites next to the condensate by numerical simulations. We analyze the complementary cumulative distribution function CCDF $ \equiv 1- \sum_{m'=1}^{m} P(m') $. Here, $m$ represents the mass on the site adjacent to the condensate. In Fig. \ref{subfig:steady state}, the main plot shows the scaled CCDF plotted as CCDF$L^2$ against $m/L^3$. We see that in steady state the CCDF $\sim m^{1-\theta}$, with $\theta=5/3$.

	The CCDF of masses relative to the condensate depicts a clear crossover as $r/L$ varies, with the exponent $\theta$ changing from $4/3$ to $5/3$ (see Fig.~\ref{fig:crossover}). When \textcolor{black}{$r/L$ is small}, the $\theta=4/3$ region is small, and the $\theta=5/3$ region dominates. As $r/L$ increases, the $\theta=4/3$ region continues to expand until it prevails everywhere. \textcolor{black}{Thus, a \textit{macroscopic} number of sites exhibit a power-law $P(m)\sim m^{-5/3}$ in steady state. We remark that the same power law  has been found in the case of the charge-injection model~\cite{Takayasu_1989}.
		When  positive and negative charges are input with equal probability, we have $\langle I \rangle = 0$. Then the leading behavior of $p(m)$ with $m$, where $m$ now represents the net charge at any site, is determined by the $\langle I^2 \rangle$ term in Eq.~(\ref{expansionphi}), yielding the power-law $p(m) \sim m^{-5/3}$ to leading order in $m$.
		  For mass injection, we have $\langle I \rangle \ne 0$; however, the mass distribution in a finite fraction of the sites still shows the power-law $P(m) \sim m^{-5/3}$. It would be interesting if a connection could be drawn between the two cases. }

\begin{figure}
	\subfloat[\label{subfig:mass1}]{
		\includegraphics[width=0.85\columnwidth]{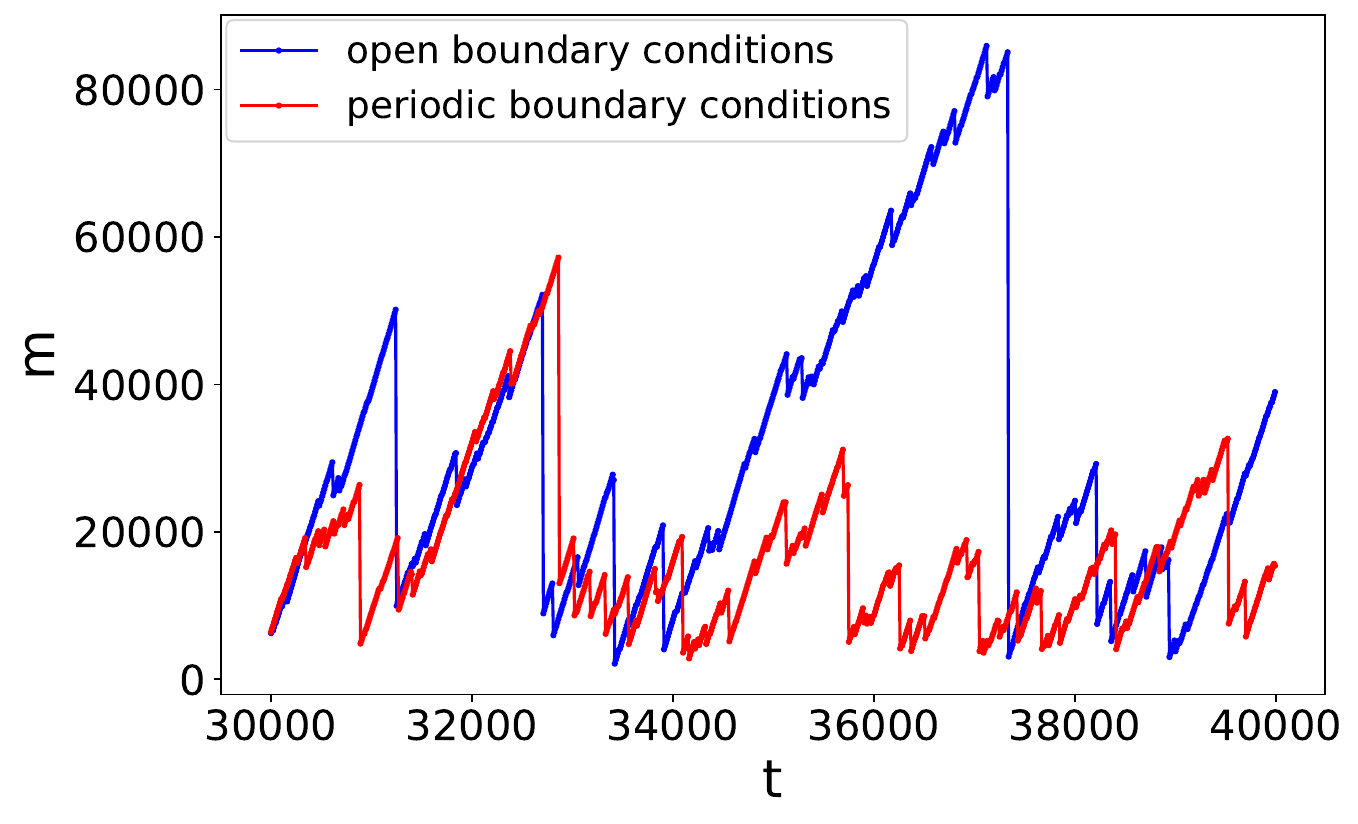}
	}
	\label{fig:mass1}
	\subfloat[\label{subfig:mass2}]{
		\includegraphics[width=0.85\columnwidth]{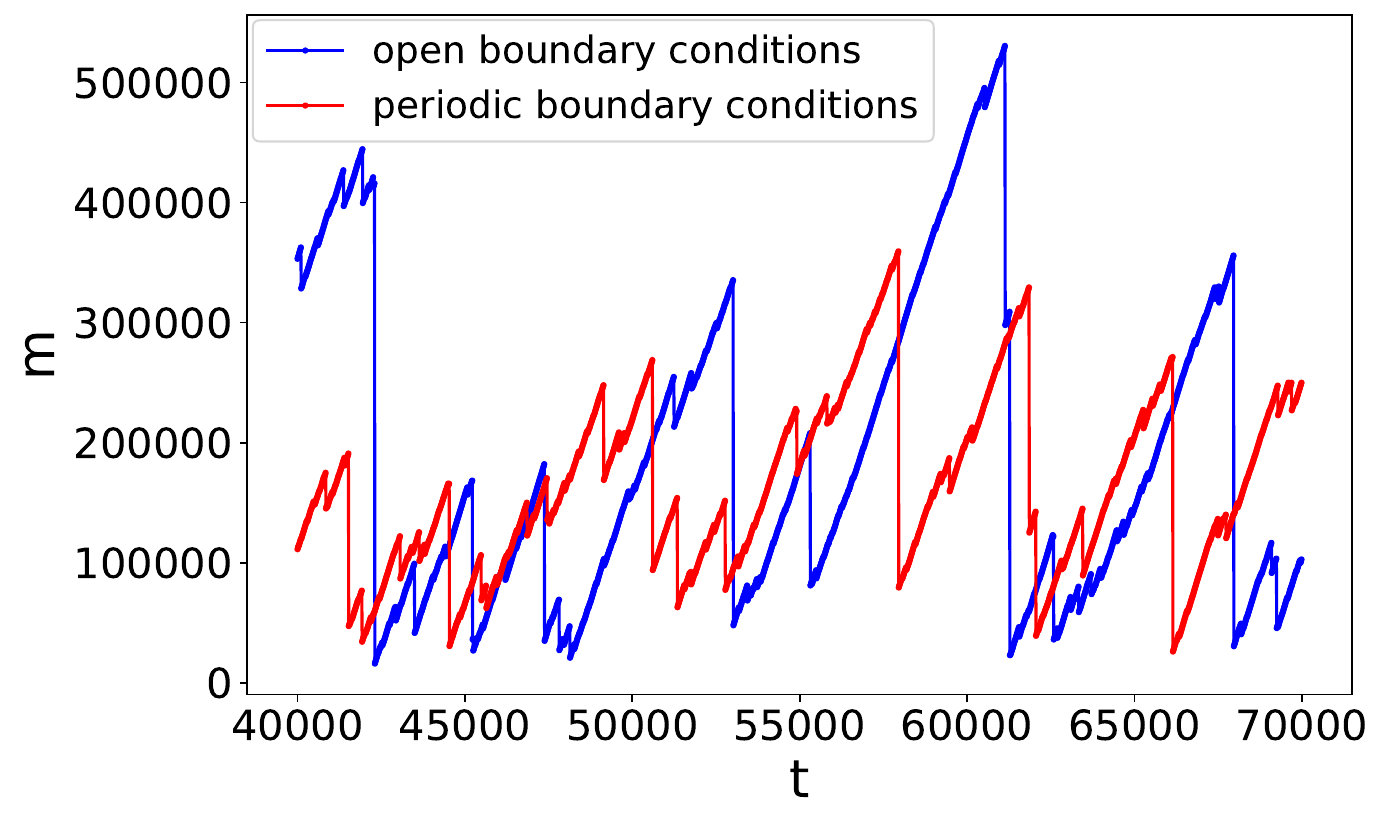}
	}
	\label{fig:mass2}
	\caption{(a) Plot of $M(t)$ vs $t$ for open boundary conditions (blue lines), and $M_{rest}(t)$ vs $t$ for periodic boundary conditions (red lines) for $L=50$ and for (b) $L=100$.  }
	
	\label{fig:mass}
\end{figure}
	
	These steady state results also have interesting implications for the regime $t\ll t^*$.Due to diffusive correlations, the system can be thought of as comprised of subsystems of size $\mathcal{L}(t)$. This length scale effectively replaces the system size $L$ in this regime. The main plot of Fig. \ref{subfig:cumulative} shows the CCDF of masses next to the condensate scaled using $\mathcal{L}(t)$ in place of $L$. The inset in Fig. \ref{subfig:cumulative} shows the unscaled CCDFs in the bulk and adjacent to the condensate at different times, highlighting the presence of two distinct power law behaviors.
	
		\begin{figure*}[htbp]
		\hspace{-1.0cm}
		\subfloat[\label{fig:S2}]{
			\includegraphics[width=0.75\columnwidth]{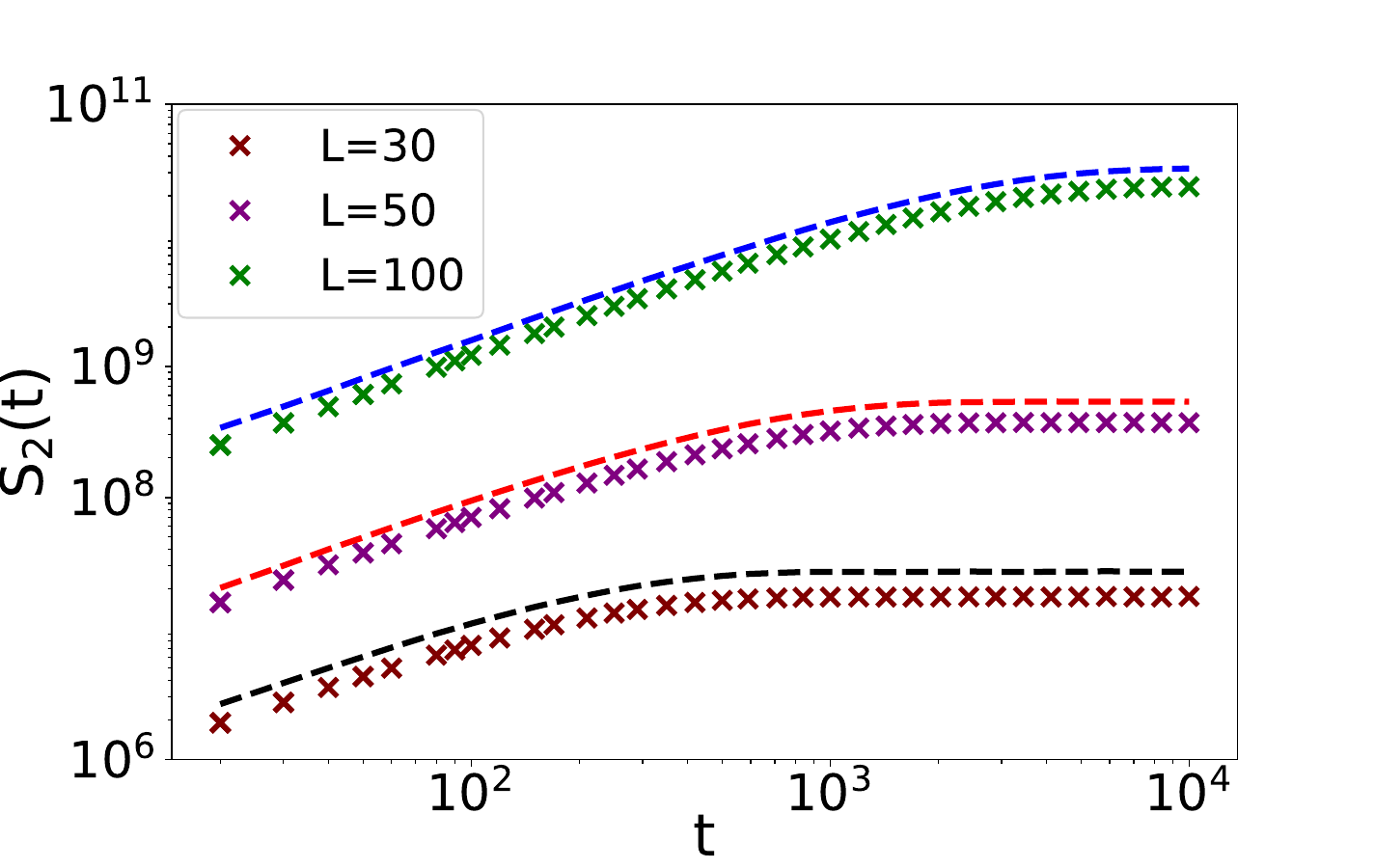}
		}\hspace{-0.89cm}
		\subfloat[\label{fig:S4}]{
			\includegraphics[width=0.75\columnwidth]{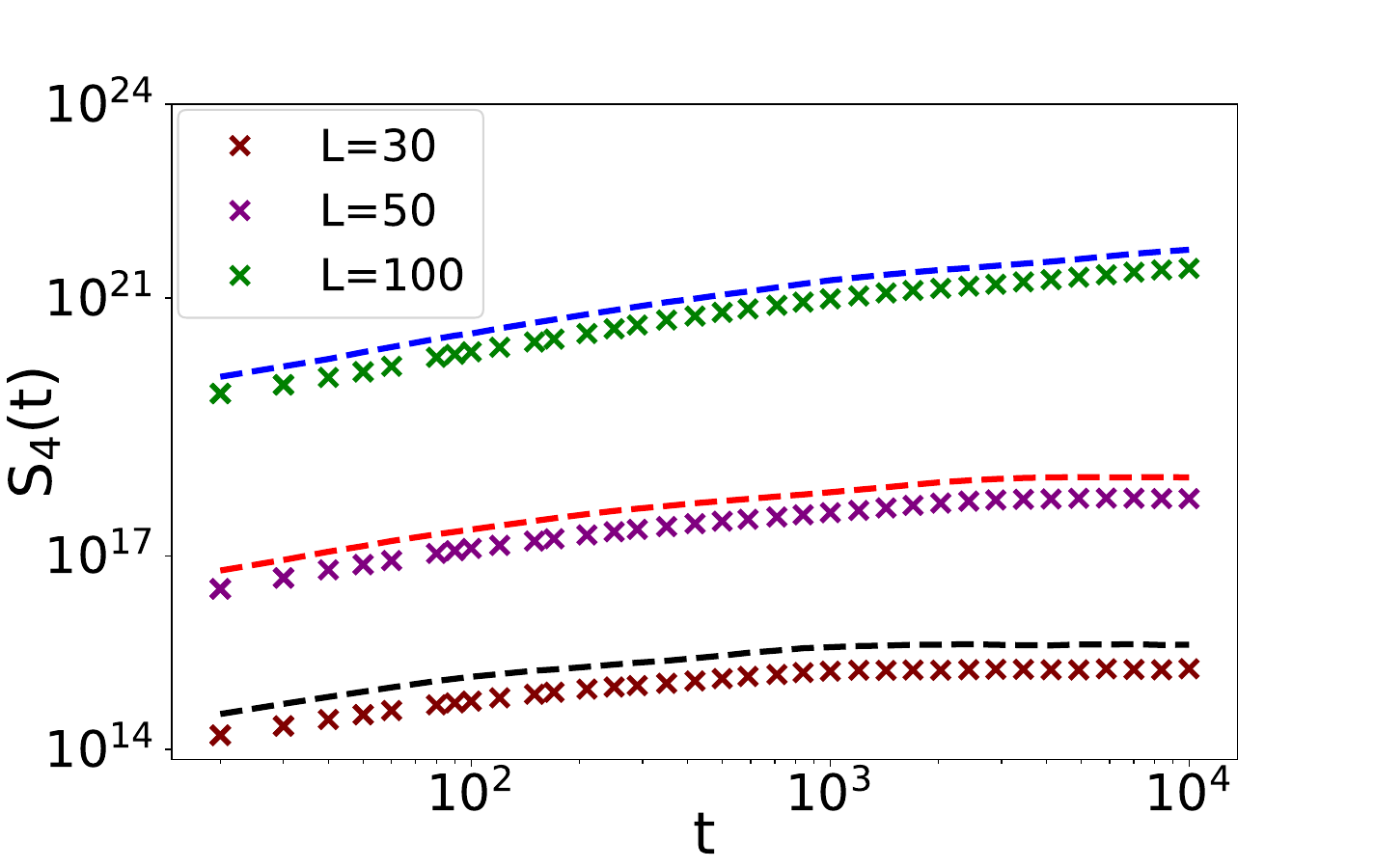}
		}\hspace{-0.97cm}
		\subfloat[\label{fig:inter}]{
			\includegraphics[width=0.75\columnwidth]{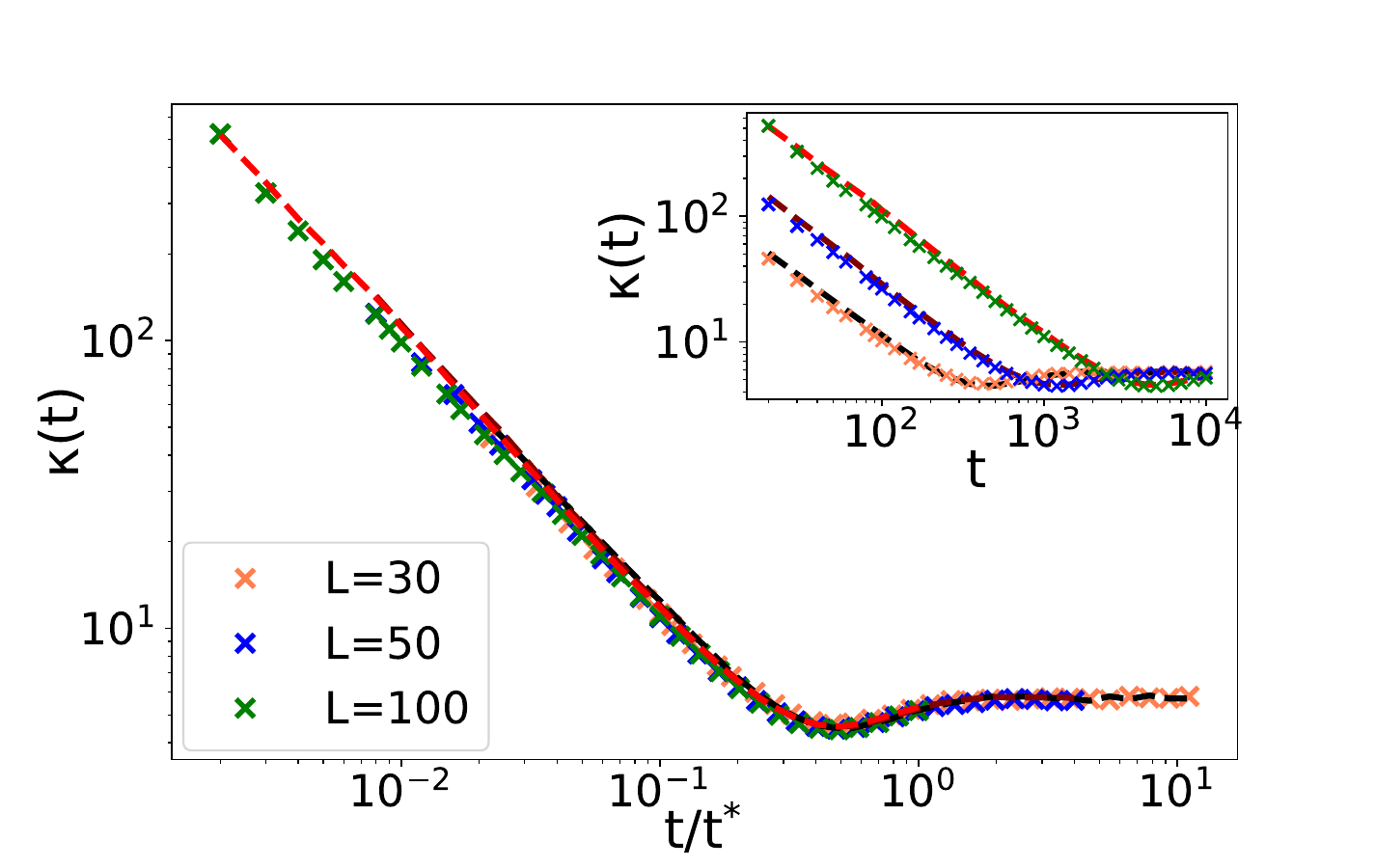}
		}\hspace{-0.95cm}
		\caption{Structure functions: (a) $S_2(t)$ and (b) $S_4(t)$ plotted against $t$ for system sizes, namely, $L=30$, $50$ and $100$. The dashed lines represent system with open boundary conditions and the crosses correspond to periodic boundary conditions. (c)  Scaling collapse for $\kappa(t)$ when plotted against $t/t^*$ for $L=30$, $L=50$ and, $L=100$. The dashed lines and the crosses  correspond to  open and periodic boundary conditions, respectively. Inset: Unscaled plots.  }
	\end{figure*}
	
	In fact, the two powers $\theta=4/3$ and $5/3$ show up prominently even in a \textit{single} configuration of a large system for $t\ll t^*$. To demonstrate this, we simulated a system with $L=10^7$. We identified the largest mass in each subsystem of length \(\mathcal{L}(t)= t^{1/2}\) as a condensate. The number of such subsystems $N = L/\mathcal{L}(t)$. Then, for each time we obtained frequency distributions: (1) for masses on the sites next to the condensates and (2) for masses across the full system, which we refer to as the bulk. We thereby found that the sites next to the condensates exhibit the mass distribution $P(m) \sim m^{-5/3}$. As time progresses, the size of each subsystem increases and $N$ decreases. However, since the mass distribution around the condensate depends only on the ratio $r/\mathcal{L}(t)$, the region influenced by the condensate grows proportional to $\mathcal{L}(t)$. As a consequence, a $\textit{finite fraction}$ of the system is expected to display a different power law from that of the bulk due to the presence of condensates, even though the entire system remains translationally invariant.

	\section{Intermittency}

	Finally, we point out that in the steady state of Takayasu model with open boundary conditions, the exits of condensates with mass $\sim O(L^3)$, lead to large crashes of the total mass $M(t)$ of the system in time (see Fig. \ref{fig:mass}). The crashes are followed by a build-up of $M(t)$ due to input, followed again by crashes, and so on in an intermittent fashion, reminiscent of dragon king events \cite{sornette2012dragon,rdsouza}.

	Intermittency, which signals breakdown of self-similarity, is quantified by the behavior of the structure functions $S_n(t)=\left<[M(t)-M(0)]^n\right> $ where $\left<....\right>$ denotes average over histories \cite{frisch1995,sachdeva2013}. A quantitative measure is provided by the flatness $\kappa(t)= S_4(t)/S_2(t)^{2}$, studied earlier in aggregation-fragmentation models with influx and outflux at the boundaries \cite{sachdeva2013}. For self-similar signals, $S_n(t) \sim t^{\gamma n}$ as $t/t^* \rightarrow 0$, where $\gamma$ is a constant. For intermittent signals, structure functions deviate from this form \cite{frisch1995,sachdeva2013,sachdeva2014}. We find that both $S_2(t)$ and $S_4(t) \sim t^{\phi}$ (see Figs. \ref{fig:S2} and \ref{fig:S4}), and therefore $\kappa \sim (t/t^*)^{-\phi}$ with $\phi \simeq 0.9$ (see Fig. \ref{fig:inter}). This signals strong intermittency.

	Interestingly, intermittency shows up even with periodic boundary conditions if we focus on $M_{rest}(t)$, the mass in the system excluding the condensate. $M_{rest}(t)$ has fixed mean value in steady state, but shows intermittent fluctuations (see Fig. \ref{fig:mass}) with a divergent $\kappa(t)$, as the condensate acts as an absorbing boundary. This is clearly illustrated in Fig. \ref{fig:inter}.

	\section{Conclusion}
	
\textcolor{black}{We have established, with the help of  exact results for the mass distribution  $p(m,t)$ for finite $L$ and $t$, that in the Takayasu mass model there are condensates that lead to a hump in the distribution when $t\ll t^*=L^2$ and to a sharper peak when $t\gg t^*$. We attribute the occurrence of condensates to the maximal term effect, whereby the maximal term in a slow power-law decay carries a finite fraction of the sum of the terms.} In steady state, the mass profile is strongly influenced by the condensate.  \textcolor{black}{Interestingly, the statistical properties of the system are stationary when viewed from a frame co-moving with the condensate.} An unexpected outcome is the occurrence of a new power law $P(m) \sim m^{-5/3}$ in a macroscopic region close to it.
 Finally, successive build-ups and exits of the condensates from the Takayasu model with open boundaries lead to crashes of the total mass, characterized by strong intermittency.
	
	In a broader vein, it would be interesting to look for strong imprints of condensates on the static and dynamic properties of other systems obeying aggregation-diffusion dynamics. On the theoretical end, the conserved mass aggregation model \cite{majumdar1998nonequilibrium,majumdar2000nonequilibrium} and the in-out model \cite{Skrishnamurthy_2000,rajesh2004nonequilibrium} both exhibit condensates, and  are therefore expected to share some of the signatures discussed above. Experimentally, biomolecular aggregates that are described within the Takayasu  framework of aggregation and input \cite{brangwynne_2011,brangwynne_2018,brangwynne_2023}, would be promising candidates to look for these signatures.
	\\
	\section{Acknowldegement}
	We thank Arghya Das and Himani Sachdeva for helpful discussions, and Deepak Dhar for a critical reading of the manuscript. We also acknowledge Eli Barkai and Punyabrata Pradhan for their useful comments. M.B. acknowledges the support of the Indian National Science Academy (INSA). We acknowledge the support of the Department of Atomic Energy, Government of India, under Project Identification No. RTI4007. 

\bibliography{references}
\end{document}